\documentclass[12pt]{article}
 \usepackage{epsfig}
 \def\be{\begin{equation}}
 \def\ee{\end{equation}}
 \def\bea{\begin{eqnarray}}
 \def\eea{\end{eqnarray}}
 \usepackage{graphicx}

 \catcode`\@=11
 \def\lsim{\mathrel{\mathpalette\@versim<}}
 \def\gsim{\mathrel{\mathpalette\@versim>}}
 \def\@versim#1#2{\vcenter{\offinterlineskip
 \ialign{$\m@th#1\hfil##\hfil$\crcr#2\crcr\sim\crcr } }}
 \catcode`\@=12

 \catcode`@=12
\begin{document}
 \thispagestyle{empty}
 \begin{flushright}
 UCRHEP-T630\\
 Mar 2025\
 \end{flushright}
 \vspace{0.6in}
 \begin{center}
 {\Large \bf Scotogenic Peccei-Quinn and One-Higgs Quark \\
and Lepton Model with Flavor Symmetry\\}
 \vspace{1.5in}
 {\bf Ernest Ma\\}
 \vspace{0.1in}
{\sl Department of Physics and Astronomy,\\ 
University of California, Riverside, California 92521, USA\\}
 \vspace{1.2in}

\end{center}

\begin{abstract}\
Implementing the Peccei-Quinn mechanism to enforce $CP$ conservation in 
strong interactions with heavy dark quarks, a scotogenic model of quark 
masses with $Z_3 \times Z_3$ symmetry is discussed where there is only 
one Higgs doublet. A natural extension to the lepton sector using 
$A_4 \times A_4$ is also presented.
\end{abstract}

\newpage

\baselineskip 24pt
\noindent \underline{\it Introduction}~:~ 
The most well-known and natural solution to the appearance of a $CP$ 
nonconserving phase in quantum chromodynamics is the Peccei-Quinn (PQ) 
mechanism~\cite{pq} based on an anomalous $U(1)$ global symmetry.  It results 
in an axion~\cite{kc10} with a very small mass which may be a component of 
the dark matter of the Universe.  There is however another consequence as 
pointed out in Ref.~\cite{dmt14}.  A residual exactly conserved $Z_2$ 
symmetry always exists.  Whereas it usually corresponds merely to 
$(-1)^{3B}$ and ensures the stability of the proton, it is also possible 
that it is in fact a dark parity under which the lightest odd particle is 
a component of dark matter in addition to the axion.

In this paper, PQ charges are assigned to heavy quarks as well as scalar 
flavons carrying $Z_3 \times Z_3$ symmetry, so that first and second 
families of 
quarks obtain scotogenic masses~\cite{m06,t96}. For previous related studies 
involving PQ symmetry, see Refs.~\cite{dmt14,m15,mot17,mrz18,mt18}

\noindent \underline{\it Model}~:~
Quarks of the standard model (SM) have no PQ charge.  They do transform 
under $Z_3 \times Z_3$ as shown in Table~1, where $\omega^3=1$.
\begin{table}[tbh]
\centering
\begin{tabular}{|c|c|c|}
\hline
quarks & $Z_3^L$ & $Z_3^R$ \\ 
\hline
$(t,b)_L$ & 1 & 1 \\ 
$(c,s)_L$ & $\omega$ & 1 \\ 
$(u,d)_L$ & $\omega^2$ & 1 \\ 
\hline
$t_R,b_R$ & 1 & 1 \\ 
$c_R,s_R$ & 1 & $\omega$ \\
$u_R,d_R$ & 1 & $\omega^2$ \\ 
\hline
\end{tabular}
\caption{PQ and $Z_3 \times Z_3$ assignments of quarks.}
\end{table}
Hence only $m_t$ and $m_b$ are tree-level masses coming from the one 
Higgs doublet $\Phi=(\phi^+,\phi^0)$.  

The new particles are listed in Table~2. A Majorana color fermion octet $\Psi$ 
with unit PQ charge generates the axion. 
\begin{table}[tbh]
\centering
\begin{tabular}{|c|c|c|c|c|c|c|}
\hline
particle & $SU(3)_C$ & $SU(2)_L$ & $U(1)_Y$ & PQ & $Z_3^L$ & $Z_3^R$ \\ 
\hline
$\Psi$ & 8 & 1 & 0 & 1 & 1 & 1 \\ 
$(a,v)_L$ & 3 & 2 & 1/6 & $1$ & 1 & 1 \\ 
$(a,v)_R$ & 3 & 2 & 1/6 & $-3$ & 1 & 1 \\ 
$a'_R,v'_R$ & 3 & 1 & $2/3,-1/3$ & $1$ & 1 & 1 \\ 
$a'_L,v'_L$ & 3 & 1 & $2/3,-1/3$ & $-3$ & 1 & 1 \\ 
\hline
$\sigma$ & 1 & 1 & 0 & 2 & 1 & 1 \\ 
$\zeta$ & 1 & 1 & 0 & 4 & 1 & 1 \\
$\kappa$ & ! & 1 & 0 & 6 & 1 & 1 \\ 
\hline
$\eta_0$ & 1 & 1 & 0 & 3 & 1 & 1 \\ 
$\eta_1^L$ & 1 & 1 & 0 & 3 & $\omega$ & 1 \\ 
$\eta_2^L$ & 1 & 1 & 0 & 3 & $\omega^2$ & 1 \\ 
$\eta_1^R$ & 1 & 1 & 0 & 3 & 1 & $\omega$ \\ 
$\eta_2^R$ & 1 & 1 & 0 & 3 & 1 & $\omega^2$ \\ 
\hline
\end{tabular}
\caption{PQ and $Z_3 \times Z_3$ assignments of new particles.}
\end{table}
The scalar potential consisting of $\sigma,\zeta,\kappa$ is given by
\begin{eqnarray}
V_0 &=& -m^2_\sigma \sigma^*\sigma + m^2_\zeta \zeta^*\zeta + m^2_\kappa \kappa^* 
\kappa + {1 \over 2} \lambda_\sigma (\sigma^*\sigma)^2 + {1 \over 2} 
\lambda_\zeta (\zeta^*\zeta)^2 + {1 \over 2} \lambda_\kappa (\kappa^*\kappa)^2 
\nonumber \\ &+& \lambda_{\sigma\zeta} (\sigma^*\sigma)(\zeta^*\zeta) + 
\lambda_{\sigma\kappa} (\sigma^*\sigma)(\kappa^*\kappa) + \lambda_{\zeta\kappa} 
(\zeta^*\zeta)(\kappa^*\kappa) + [\mu_\zeta \zeta^*\sigma\sigma + \mu_\sigma 
\kappa^*\zeta\sigma + h.c.],
\end{eqnarray}
where the would-be allowed terms
\begin{equation}
\kappa^*\sigma\sigma\sigma, ~~~ \kappa^*\sigma^*\zeta \zeta
\end{equation}
are forbidden by the imposition of a specific $Z_4$ symmetry under which 
$\sigma \sim -1, \Psi \sim i$, with all other particles $\sim 1$.  It is 
preserved by all dimension-4 terms, but broken softly by the dimension-3 
term $\mu_\sigma$.  With large positive $m^2_\sigma,m^2_\zeta,m^2_\kappa$ of 
the same order of magnitude, it is easily shown that
\begin{equation}
\langle \sigma \rangle \simeq \sqrt{m^2_\sigma \over \lambda_\sigma}, ~~~ 
\langle \zeta \rangle \simeq {-\mu_\zeta \langle \sigma \rangle^2 \over 
m^2_\zeta + \lambda_{\sigma\zeta} \langle \sigma \rangle^2}, ~~~ 
\langle \kappa \rangle \simeq {-\mu_\sigma \langle \sigma \rangle 
\langle \zeta \rangle \over m^2_\kappa + \lambda_{\sigma\kappa} 
\langle \sigma \rangle^2}.
\end{equation}
Now $\mu_\sigma$ is already understood to be small beacuse it breaks the 
specific $Z_4$ symmetry mentioned earlier.  In addition, there is a universal 
$Z_4$ symmetry which applies to all particles~\cite{m25-1,m25-2}, under which
\begin{equation}
A_\mu \sim 1, ~~~ \phi \sim -1, ~~~ \psi_L \sim i, ~~~ \psi_R \sim -i,
\end{equation}
where $A_\mu$ is any vector gauge boson, $\phi$ any scalar, $\psi_L$ any 
left-handed fermion, $\psi_R$ any right-handed fermion.  Since the $\mu_\zeta$ 
term breaks this $Z_4$ softly and explicitly. it may also be assumed 
naturally small.  As a result, 
$\langle \kappa \rangle << \langle \zeta \rangle << \langle \sigma \rangle$.

The axion scale is determined by $\langle \sigma \rangle$ which should be of 
order $10^9$ to $10^{12}$ GeV from astrophysical constraints. 
From the Yukawa coupling $\sigma^* \Psi \Psi$, $m_\Psi$ is large, whereas the 
dark quarks obtain their masses of a few TeV from $\langle \zeta \rangle$, 
which may be of order $10^7$ GeV.  With the addition of the dark flavons 
$\eta_i$, the terms $\kappa^*\eta_i\eta_j$ are also small because 
$\langle \kappa \rangle$ may be of order TeV.  Note that the undesirable 
terms $\sigma^*\zeta^*\eta_i\eta_j$ are forbidden by the specific $Z_4$ 
symmetry assumed earlier.

\noindent \underline{\it Scotogenic Quark Masses}~:~
Using the new particles of Table~2, radiative quark masses are obtained as 
shown in Fig.~1.
\begin{figure}[htb]
\vspace* {-3.5cm}
\hspace*{-3cm}
\includegraphics[scale=1.0]{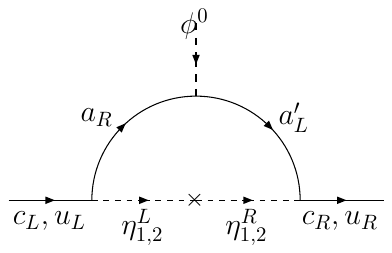}
\vspace{-23.0cm}
\caption{Scotogenic Quark Masses.}
\end{figure}
The analogous diagrams linking $s_L,d_L$ to $s_R,d_R$ through $v_R,v'_L$ 
are not shown but should be obvious. There are also diagrams linking $t_{L,R}$ 
and $b_{L,R}$ to the other quarks through $\eta_0$. 
If $Z_3 \times Z_3$ is exact, the $5 \times 5$ mass-squared matrix 
${\cal M}^2_\eta$ is diagonal, and all such amplitudes are zero. 
Allowing soft $Z_3 \times Z_3$ breaking in the form $\eta_i^* \eta_j$ for 
$i \neq j$, so that 
\begin{equation}
{\cal M}^2_\eta = U^\dagger {\cal M}^2_{diag} U,
\end{equation}
the link between $\eta_i$ and $\eta_j$ in Fig.~1 is then given by 
$\sum_k U^\dagger_{ik} m_k^{-2} U_{kj}$, where $m_k$ are the new $\eta$ mass 
eigenvalues.  For $i \neq j$, $\sum_k U^\dagger_{ik} U_{kj} = 0$ which ensures 
the one-loop finiteness of Fig.~1.  The $2 \times 2$ mass matrix linking 
$(\bar{a}_L \bar{a}'_L)$ to $(a_R,a'_R)$ is 
\begin{equation}
{\cal M}_a = \pmatrix{M_a & m_a \cr m'_a & M'_a},
\end{equation}
where $M_a,M'_a$ come from $\langle \zeta \rangle$, $m_a$ from 
$\langle \bar{\phi}^0 \rangle$, and $m'_a$ from $\langle {\phi}^0 \rangle$. 
It is clear that a $Z_2$ symmetry remains for all the internal particles 
of the loop in Fig.~1.

Under $Z_3^L \times Z_3^R$, only the $\bar{t}_L t_R$ and $\bar{b}_L b_R$ 
entries in the two respective $3 \times 3$ ${\cal M}_U$ and ${\cal M}_d$ 
mass matrices couple directly to the one Higgs doublet.  The other entries 
depend on the soft breaking $\eta^*_i \eta_j$ terms.  Suppose they obey the 
exchange symmetry $\eta^L_i \leftrightarrow \eta^R_i$, then both ${\cal M}_u$ 
and ${\cal M}_d$ are Hermitian, which is the starting arbitrary assumption 
of many studies.  Here it is justified in the context of an exchange 
symmetry in the soft breaking flavon sector.  Parallel ${\cal M}_u$ and 
${\cal M}_d$ structures are also guaranteed, including texture zeros, such 
as in the Fritzsch ansatz. 

\noindent \underline{\it Observable Consequence}~:~
As pointed out in Ref.~\cite{fm14}, there are now anomalous one-loop Higgs 
interactions. The Yukawa coupling $h \bar{q} q$ is no longer $m_q/v\sqrt{2}$ 
as in the SM. There may also be nondiagonal couplings such as 
$h \bar{t}_L c_R$, etc.  If $m_a,m'_a << M_a,M'_a$ in Eq.~(6), the Higgs 
couplings of Fig.~1 are proportional to the corresponding mass entries 
as in the SM.  However, there are now additional Higgs couplings, such as 
$\lambda_{01} (v + h/\sqrt{2})^2 \eta^*_1 \eta_1$.  This implies an anomalous 
$h \bar{t}_L c_R$ coupling given by $\sqrt{2}v \lambda_{01} m_{tc}/m^2_{eff}$, 
where $m_{eff}$ is an effective combination of $\eta$ masses.  For a 
study of the possible observation of $t \to h c$ decay at the Large Hadron 
Collider (LHC), see Ref.~\cite{gjk21}.

\noindent \underline{\it Lepton Sector}~:~
\begin{table}[tbh]
\centering
\begin{tabular}{|c|c|c|c|}
\hline
particle & PQ & $A_4^L$ & $A_4^R$ \\ 
\hline
$(N,E)_L$ & $1$ & 1 & 1 \\ 
$(N,E)_R$ & $-3$ & 1 & 1 \\ 
$E'_R$ & $1$ & 1 & 1 \\ 
$E'_L$ & $-3$ & 1 & 1 \\ 
$S_L$ & $-3$ & 1 & 1 \\
\hline
$\eta^L_{3,4,5}$ & 3 & 3 & 1 \\ 
\hline
\end{tabular}
\caption{PQ and $A_4 \times A_4$ assignments of new particles.}
\end{table}
Since $Z_3$ is a subgroup of $A_4$, $Z_3 \times Z_3$ may be replaced by 
$A_4 \times A_4$, which fits nicely to the $A_4$ model of leptons proposed 
recently~\cite{m25} with some modifications.  The three lepton doublets 
transform as $\underline{3}$ under $A_4^L$, whereas the singlets transform as 
$\underline{1},\underline{1}',\underline{1}''$ under $A_4^R$ which are the 
same as $1,\omega,\omega^2$.  The additional particles are shown in Table~3.

The resulting radiative charged lepton masses are shown in Fig.~2.
\begin{figure}[htb]
\vspace* {-3.5cm}
\hspace*{-3cm}
\includegraphics[scale=1.0]{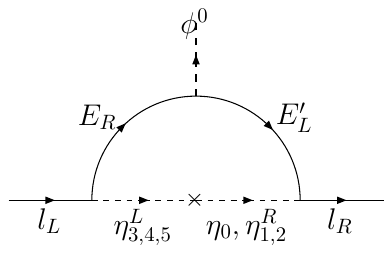}
\vspace{-23.0cm}
\caption{Scotogenic Charged Lepton Masses.}
\end{figure}
The soft breaking which mixes $\eta^L_{3,4,5}$ with $\eta_0,\eta^R_{1,2}$ 
is of the form $U_\omega$ times a diagonal matrix, where
\begin{equation}
U_\omega = {1 \over \sqrt{3}} \pmatrix{1 & 1 & 1 \cr 1 & \omega & \omega^2 \cr 
1 & \omega^2 & \omega},
\end{equation}
as explained in Ref.~\cite{m25}.  The neutrino masses are determined by 
Fig.~3.
\begin{figure}[htb]
\vspace* {-5.0cm}
\hspace*{-3cm}
\includegraphics[scale=1.0]{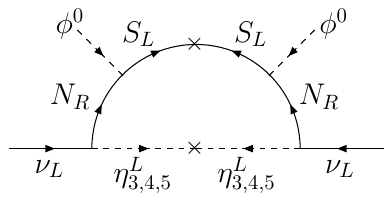}
\vspace{-21.5cm}
\caption{Scotogenic Neutrino Masses.}
\end{figure}
The difference here is that $S$ is not ultralight because of the 
$\kappa SS$ term. 

\noindent \underline{\it Dark Matter}~:~
Depending on the exact value of $\langle \sigma \rangle$, the axion may 
contribute only to part of the dark matter density of the Universe. 
The rest is made up of the lightest scalar flavon in this scenario.  It 
may be produced by $hh$ annihilation, but its couplings to $u$ and $d$ quarks 
may be suppressed so that it is not easily detected at underground experiments. 
For a comprehensive study, see Ref.~\cite{GAMBIT17}.  Its mass should again 
be of order a few TeV.

\noindent \underline{\it Concluding Remarks}~:~ 
The SM may be complete as it is, with just one Higgs doublet.  To answer other 
questions, such as strong CP conservation and the flavor structure of quarks 
and leptons as well as the origin of neutrino mass, a dark sector based on 
PQ symmetry is an idea worth exploring.  In addition to the axion, a 
residual $Z_2$ symmetry supplies another dark matter candidate.  A 
$Z_3 \times Z_3$ model of radiative quark masses is discussed here with 
desirable properties.  An extension to the lepton sector is accomplished 
with $A_4 \times A_4$.

\noindent \underline{\it Acknowledgement}~:~
This work was supported in part by the U.~S.~Department of Energy Grant 
No. DE-SC0008541.  

\baselineskip 18pt
\bibliographystyle{unsrt}

\end {document}